New method to estimate stability of chelate complexes

Running title: Stability of chelate complexes

F.V.Grigoriev[a,b,*] , A.Yu.Golovacheva[b,c] , A. N. Romanov[a,b], O.A. Kondakova[a,b] ,V. B. Sulimov[a,b,c].

*[a] Research Computing Center, Lomonosov Moscow State University, Moscow, Russia; [b] Victory Pharmaceutical, Troitsk, Russia; [c] Dimonta,Ltd., Moscow, Russia.*

A new method allowing calculation of the stability of chelate complexes with $Mg^{2+}$ ion in water have been developed. The method is based on two-stage scheme for the complex formation. The first stage is the ligand transfer from an arbitrary point of the solution to the second solvation shell of the $Mg^{2+}$ ion. At this stage the ligand is considered as a charged or neutral rigid body. The second stage takes into account disruption of coordinate bonds between $Mg^{2+}$ and water molecules from the first solvation shell and formation of the bonds between the ligand and the $Mg^{2+}$ ion. This effect is considered using the quantum chemical modeling. It has been revealed that the main contribution to the free energy of the complex formation $\Delta G_b$ is caused by the disruption/formation of the coordinate bonds between $Mg^{2+}$, water molecules and the ligand. Another important contribution to the complex formation energy is change of electrostatic interactions in water solvent upon the ligand binding with $Mg^{2+}$ ion. For all complexes under investigation the reasonable (in frame of 2 kcal/mol) agreement between calculated and experimental $\Delta G_b$ values are achieved.

## Introduction

Ability of organic ligands to form chelate complexes with metal ions should be taken into consideration for such processes as design of metalloprotein inhibitors, determination of structure of guest-host complex stability and some others. Ability of a ligand to form chelate complexes is quantitatively characterized by the free energy of complex formation. Since the complex formation is accompanied by coordination bond breakage/formation, a method for the $\Delta G_b$ calculation should include the quantum chemical modeling. Currently, some publications on modeling geometric and thermodynamic characteristics of metal cations and their complexes with small molecules dissolved in water are available. The experimental data can be found in [1–9].

Study of water molecules sequential association with $Mg^{2+}$, $Ca^{2+}$, $Be^{2+}$ и $Zn^{2+}$ cations in frame of the density functional theory with hybrid functional B3LYP using LANL2DZ and 6-311+G(2d,2p) basis sets is presented in [10]. As for the $Mg^{2+}$ ion, the calculated length of coordination bonds in clusters of the metal ion and water molecules is in an agreement with experiments within 0.005 Å and 0.01 Å for the first and second coordinating sphere, respectively. In the case of the $Ca^{2+}$ ion the basis set variation comparatively weakly influences on the ion-water binding energy. Authors of [11] investigated singly and doubly charged ion complexes with water, formaldehyde and ammonia using DFT/B3LYP, MP2 and CCSD(T) theory level and 6-311++G(3df,3pd) and smaller basis sets. It was found that the change in the basis set from 6-311G(d,p) to 6-311++G(3df,3pd) decreases the $Mg^{2+}$-O bond length less than 0.03 Å. Comparison of experimental and calculated enthalpy of the water molecules binding to $Li^{+}$, $K^{+}$, $Na^{+}$ cations, $\Delta H^{298}$, was performed in [11]. The best agreement with experiment was obtained at MP2/6-311++G(3df,3pd) and B3LYP/6-311++G(3df,3pd) theory levels. The B3LYP level underestimates $\Delta H^{298}$ value by 2 kcal/mol for $Na^{+}$ cation. CCSD(T) method with this basis sets systematically underestimates $\Delta H^{298}$ by 2÷3 kcal/mol.

[*] Corresponding author. Email: fedor.grigoriev@gmail.com

Quantum chemical simulations of complex formation between small molecules and metal cations were carried out in [12-14]. Vacuum properties of $Li^{+}$, $Na^{+}$ и $Mg^{2+}$ ion complexes with formaldehyde, formic acid, formate anion and formamide were calculated using B3LYP/6-31G* and MP2 levels of theory in [12]. The complex formation energy calculated using MP2 is higher by ~4-6% than one calculated at B3LYP level. Geometry of complexes were close to experimental ones. Thermodynamic values of the complex formation of the small organic molecules (including phosphoranes) with one or two magnesium ion and geometry of these complexes were calculated in [13]. The calculations were performed within the DFT/6-311++G(3df,2p) theory level in gas phase. The continuum (implicit) solvation models, PCM (Polarizable Continuum Model) and COSMO (COnductor –like Screening MOdel) [15] for water molecules in the outer sphere were used. The complex formation energy is larger for bidentate complexes than for monodentate ones, and negatively charged ligands form stronger bonds with the magnesium ion comparatively with neutral ones. It should be noted that the difference between $\Delta G_b$ values calculated with PCM и COSMO approximations reaches 10 kcal/mol. The authors explain this disagreement by shortcomings of the implicit water models.

The original procedure of complex formation in water was proposed in [14]. The first stage is ligand transfer from infinity to the second coordinating sphere of the metal cation. The second stage is formation of coordination bonds between the ligand and the cation, which is accompanied by the water molecules displacement from the inner sphere of the cation to the outer one. The quantum chemical modeling was used only at the second stage of calculations. This model results in decreasing errors caused by the implicit water model. This procedure was applied for investigation of zinc ion complexes. The geometry optimization was carried out using HF approach with an original basis set. The electron correlation was taken into account within the framework of MP2 for the geometry optimized at the HF level. The discrepancy between calculated and experimental data of $\Delta G_b$ for this two-stage scheme was within 1 kcal/mol. This discrepancy increases to 2-3 kcal/mol if the calculations were performed in the frame of the traditional approach when the free energy change during complex formation is calculated in the framework of PCM model as $G_{PCM}(AB) - G_{PCM}(B) - G_{PCM}(A)$, where $G_{PCM}(AB)$, $G_{PCM}(B)$, $G_{PCM}(A)$ are free energy of interaction of water and complex *AB*, and its components *B* and *A*, respectively.

In the present work we report the results of the molecular modeling investigation of thermodynamics of chelate complexes formation. We calculated the free energy of complex formation $\Delta G_b$ for a number of complexes of organic molecules having different size, charge and type of coordination bonds with $Mg^{2+}$ ions. Calculations were carried out using two-stage scheme, which is derived from the scheme presented in [14]. Experimental values of $\Delta G_b$ for the complexes were taken from the NIST data base (National Institute of Standards and Technology) [16].

## Methods

### *Stages of the $Mg^{2+}L$ complex formation*

As appears from recent literature studies of a metal cation hydration and its complexes with small organic molecules, the most suitable way to take into account solvation effect is to use hybrid water models (see for example [17]). In such models the most important water molecules are considered explicitly while the others are described using implicit models like PCM. However this approach can lead to errors caused by roughness of implicit models [13-14] (see Introduction). These errors may be minimized by decomposition of the magnesium-ligand ($Mg^{2+}L$) complex formation into two stages [14]:

$$[Mg^{2+}(H_2O)_6](aq) + L(aq) \xrightarrow{(a)} [Mg^{2+}(H_2O)_6], L(aq) \xrightarrow{(b)} [Mg^{2+}L(H_2O)_{6-n}], (nH_2O)(aq) \quad (1),$$

where L – a ligand, *n* – the number of coordination bonds of the ligand with the magnesium ion, (aq) – denotes bulk water considered within the implicit water model, molecules placed out of square brackets are situated outside the first coordinating sphere (CS) of the metal cation.

At he stage (a) the ligand is transferred from an arbitrary position far from the metal cation to the second coordinating sphere of the magnesium ion. Six water molecules (first CS the $Mg^{2+}$ ion) form coordination bonds with the magnesium ion in the complex $Mg^{2+}(H_2O)_6$.

At the stage (b) the ligand initially located in the second CS of the cation (Figure 1a) substitutes *n* water molecules in the first CS (these water molecules are therefore replaced to the second CS – Figure 1b).

Free energy of the $Mg^{2+}$ complex formation with organic molecules in water $\Delta G_{b_calc}$ is calculated as

$$\Delta G_{b_calc} = \Delta G_{(a)} + \Delta G_{(b)} \quad (2),$$

where $\Delta G_{(a)}$ and $\Delta G_{(b)}$ are free energies of stages (a) and (b), respectively.

Quantum chemical calculations with PCM were used only for the second reaction stage when the ligand and a water molecule (or few water molecules in the bi- or more dentate ligand case) exchanged their positions at the first and second CSs. Changes in the shape and area of the solvent accessible surface covering all entities in the first and second CSs (the ligand, the metal cation and respective water molecules) at the stage *(b)* are small, and total charge inside the surface does not change. Within the traditional approach contribution to the change of the free energy is determined as: $G_{PCM}(\mathrm{Mg^{2+}L(H_2O)_6}) - G_{PCM}(\mathrm{Mg^{2+}(H_2O)_6}) - G_{PCM}(\mathrm{L})$. In the case of charged ligands, charges of either of the three indicated above components are different, although the shape and area of the surface around the final reaction product $\mathrm{Mg^{2+}L(H_2O)_6}$ and initial reagents $\mathrm{Mg^{2+}(H_2O)_6}$ and $G_{PCM}(\mathrm{L})$ are different too. Under existing conditions, errors caused by the use of the implicit water model increase accordingly to the scheme referred above.

The methods for the $\Delta G_{(a),(b)}$ calculation are described below in detail.

***Calculation of the free energy change at the stage (a)***

The free energy change $\Delta G_{b_calc}$ at the stage *(a)*, $\Delta G_{(a)}$, is caused by restriction of the ligand available phase space. In a number of publications [14, 18, 19] $\Delta G_{b_calc}$ is computed using Fuoss equation [20] for coupling constant of spherically symmetric ions in dielectric medium with permittivity $\varepsilon$. In order to examine the metal ion interaction with a neutral molecule, we used a more general approach. In this context Fouss equation is derived as a special case.

At this stage the ligand is considered as a rigid body having a charge or a dipole moment (for the neutral ligand). Calculating $\Delta G_{(a)}$ we are guided by the general expression for the free energy of the complex AB formation from the components A and B in the dilute solution approximation:

$$\Delta(\Delta G_{AB}) = -RT\ln\frac{VN_0\int_{AB}\exp(-\beta U_{AB})d\Gamma_{AB}}{\int\exp(-\beta U_A)d\Gamma_A\int\exp(-\beta U_B)d\Gamma_B} = -RT\ln\frac{\int_{A-B}\exp(-\beta U_{AB})d\Gamma_{A-B}}{V\exp(-\beta(U_A+U_B))} \quad (4),$$

where $U_{A(B,AB)}$ are mean force potentials, $\beta = \frac{1}{k_B T}$, the integration is performed over all coordinates of $A$ and $B$ components ($d\Gamma_A, d\Gamma_B$), coordinates of the $AB$ complex $d\Gamma_{AB}$ and relative coordinates of the $AB$ complex (within the limits where components A and B form complex) $d\Gamma_{A-B}$. The absence of dependence of $U_{A,B}$ on internal coordinates in the context of chosen approach is used in (4). Follow [20], we suppose that $U_{AB}$ is not changed at $r_{AB} < a$, where $a$ is a distance on which magnesium the cation and the ligand form the complex. Taking into account the energy of the interaction of the charges $q_A$ and $q_B$ which are placed on the distance $r_{AB}$ in the dielectric medium with permittivity $\varepsilon$, $\frac{q_A q_B}{\varepsilon r_{AB}}$, we obtain from (3) and (5):

$$\Delta G_{(a)} = -RT \ln \frac{4\pi N_0 a^3 \exp(-\frac{\beta q_A q_B}{\varepsilon a})}{3} \quad (6)$$

Hence at room temperature we have:

$$\Delta G_{(a)} = \frac{N_0 q_A q_B}{\varepsilon a} - RT \ln 2.52 * 10^{-3} a^3 = \frac{4.1 Z_A Z_B}{a} - 0.59 \ln 2.52 * 10^{-3} a^3 \quad (7) \ .$$

This expression corresponds to the coupling constant presented in [20] and used in [14, 18, 19] (the value $a$ in (7) must be in $\AA$), $Z_A, Z_B$ - charges of $A$ and $B$ components (in au).

In case of $Mg^{2+}$ ion $Z_A = +2$, and $a \cong 4$ Å (it corresponds to the distance from the center of the ligand located in the second CS to the magnesium ion) we obtain:

$$\Delta G_{(a)} = (2.1 Z_B + 1.1)\, kcal/mol \quad (8)$$

Let us consider the case of the magnesium ion complex formation with a neutral ligand having the dipole moment $d_B$. Energy of interaction of the ligand and the ion is described as:

$$W = \frac{q_A d_B \cos\alpha}{\varepsilon r^2} \quad (9).$$

Mean value of angle cosine $< \cos\alpha >= cthx - \frac{1}{x}$ is derived as result of possible orientation where $x = \frac{\beta q_A d_B}{\varepsilon r^2} = \frac{10.8 Z_A D_B}{r^2}$, $Z_A = Z_{Mg} = +2$, $D_B$ - dipole moment of molecule in $e$Å, where $e$ – the elementary charge. In view of previous case, $a \cong 4$ Å, $D_B \approx 0.5$, $x = 0.68$, $< \cos\alpha >= 0.22$. Hence

$$\Delta G_{(a)} = \frac{0.9 Z_A D_B}{a^2} - 0.59 \ln 2.52 * 10^{-3} a^3 = 1.0\ kcal/mol \quad (10) \ .$$

The main contribution to (10) is given by the second term which has entropic character while the first term in (10) depending on the dipole moment does not exceed a few hundredth of kcal/mol.

Thus, in the case of the ion interaction with a neutral ligand with a dipole moment, enthalpy component is small, and $\Delta G_{(a)} \cong 1 kcal/mol$ is determined by change in the entropy resulting from the complex formation. Dependence of $\Delta G_{(a)}$ on the dipole moment can be neglected. In the case of a negatively charged ligand, $\Delta G_{(a)}$ calculated using (8) is negative and it appreciably depends on the ligand charge.

***Calculation of the free energy change at the stage (b).***

The free energy change at the stage (b), $\Delta G_{(b)}$, resulting from the ligand transfer from the second CS to the first one is described as:

$$\Delta G_{(b)} = \Delta E_0 + \Delta G_{PCM} + \Delta G_{vib_tr} + \Delta G_{cav} + \Delta G_{non-polar} + \Delta E_{rel} \quad (11),$$

Expression (11) differs from one used for computing $\Delta G_{(b)}$ in [14] by the presence of the three additional terms - $\Delta G_{cav}, \Delta G_{non-polar}$ and $\Delta E_{rel}$ (see meaning below). Every term (with the exception for $\Delta E_{rel}$) is calculated as difference of corresponding quantities for the final and the initial products of the reaction at the stage (b). Components in expression (11) denote following.

Sum $\Delta E_0 + \Delta G_{PCM}$ is the change of the total energy of the molecular cluster containing the magnesium ion, the ligand and the six water molecule and surrounded by the solvent accessible surface (SAS) which is used in frame of the PCM approach; $\Delta E_0$ - change in the cluster internal energy, $E_0 = <\Psi|H_0|\Psi^*>$, where $\Psi$ - is the wave function of the ground state, $H_0$ - vacuum Hamiltonian; $\Delta G_{PCM}$ - polarization energy of the cluster (i.e. the energy of the electrostatic interaction of the cluster charges, placed in the atoms, and induced on the surface charges which are calculated in frame of the PCM approach); $\Delta G_{cav}$ - the change of the cavitation free energy calculated according to the procedure published in [21, 22] as $\Delta G_{cav}{}^{V} = \beta * V$, where $V$ – cavity volume in $Å^3$ bounded by SAS, $\beta = 0.0389 Å^{-3}$ kcal/mol – empirical coefficient derived from the calculation by means of the Monte-Carlo method of the cavitation free energy for a number of the different organic molecules. $\Delta G_{non-polar}$ - the change of the nonpolar part of free energy of cluster interaction with implicit water, $\Delta G_{non-polar} = -0.032 - 0.077S$ ($S$ – area of SAS in $Å^2$) [23]. $\Delta G_{vib_tr}$ - is contribution to the $\Delta G_{(b)}$ resulting from change of the oscillatory, rotational and translational degrees of freedom of molecules. Term $\Delta E_{rel}$ considers relaxation of internal geometry of water molecules located in the first CS of the magnesium ion after the ligand transfer from the second CS to infinity (stage (a)).The optimization of complexes with negatively charged ligands in the second CS showed that OH bond length of bridge water molecules that form a coordination bond with $Mg^{2+}$ ion and a hydrogen bond with the negatively charged ligand atoms exceeded the equilibrium OH bond length by several hundredth of an angstrom. Hence, the ligand transfers from infinity to the second CS leads to the relaxation of some water molecules in the first CS, and the corresponding energy effect of this relaxation is:

$$\Delta E_{rel} = E_{water}([\mathrm{Mg}^{2+}(\mathrm{H_2O})_6], \mathrm{L(aq)}) - E_{water}([\mathrm{Mg}^{2+}(\mathrm{H_2O})_6]),$$

where $E_{water}([\mathrm{Mg}^{2+}(\mathrm{H_2O})_6], \mathrm{L(aq)}), E_{water}([\mathrm{Mg}^{2+}(\mathrm{H_2O})_6])$ are internal energies of six water molecules in optimized geometry with the ligand located in the second CS and in optimized geometry without the ligand, respectively. Optimization of the geometry of the ligand does not significantly influence to the $\Delta E_{rel}$

Quantum chemical calculations were carried out using package PC GAMESS [24] with supercomputer Chebyshev of SRCC MSU [25], JSC RAS [26] and Cyberia [27].

**Results and Discussion**

Calculations of $\Delta G_{b_calc}$ were carried out for nine complexes selected from the NIST database [16]. The 2D –structure of the ligands forming the complexes with one magnesium cation $Mg^{2+}$ are shown in Figure 2. The ligands were selected accordingly to the following rules. Firstly, the NIST database does not contain an information about the geometry of the complexes; therefore the atoms of the ligands which form coordination bonds have to be determined unambiguously on the basis of their 2D- structure. Secondly, ligands have to vary in structure, size and type of the coordination bonds in order to further verify the applicability of the approach developed. Thirdly, limited computational resources for the *ab initio* quantum chemical calculations restrict the ligand size.

We used two basis sets 6-31G** and cc-pVTZ and two approaches for electron correlation: the density functional B3LYP [28-30] and the second order perturbation theory MP2 [31].

Nine complexes (see Figure 2) of small organic molecules with one magnesium cation $Mg^{2+}$ have been taken from NIST database [16] for comparison calculated binding energies with experimentally measured ones. NIST database contains the experimentally determined equilibrium constant $K = \frac{[ML]}{[M][L]}$ for the complexes formation and the protonation constant p$K_a$.

The latter can be used to determine the charge of the ligand. Experimental values of the complex formation free energy $\Delta G_{b_exp}$ were calculated $\Delta G_{b_\exp} = -RT \ln K$ (13).

The quantum chemical part of calculations for each complex contains the following stages.

1) Complete geometry optimization of the complex in vacuum.

2) Calculation of the frequencies in the optimized geometry.

3) Optimization taking into account the solvent effects using the PCM approach.

4) Qualifying calculations in optimized in stage 3) geometry.

At each stage, two calculations corresponded to the ligand in the first CS and the second CS of the magnesium ion, respectively, were carried out. At the stage 1)-3) calculations were carried out within the B3LYP/6-31G** level, at the stage 4) $\Delta E_0$ (see (11)) was also calculated using MP2 approach with cc-pVTZ basis set.

The results of the calculations of the free energy of the complex formation are listed in the Table 1.

For all the complexes except the monodentate pyridine (I) complex, the internal energy change is $\Delta E_0 < 0$ both for negatively charged and neutral ligands. It results from the chelate effect which increases the binding energy if the ligand is more than monodentate. The most substantial decrease in $\Delta E_0$ was obtained for the oxalate ion (IX), which is related to a more appreciable contribution to $\Delta E_0$ from the electrostatic energy (Z=-2) compared with the other ligands. Partly the decrease in $\Delta E_0$ is compensated by the increase in electrostatic part of the free energy of the interaction of the complex components with solvent ($\Delta G_{PCM}$ term). The contribution to the $\Delta G_{b_calc}$ from the free energy change of vibrational, rotational and translational components, $\Delta G_{vib_tr}$, is positive for all complexes. The contribution to the $\Delta G_{b_calc}$ from the free energy change of the cavitation component, $\Delta G_{cav}$, and nonpolar part of the free energy of the interaction of the complex components with solvent are relatively small and can be positive and negative both. The relaxation energy, $\Delta E_{rel}$, is positive for all complexes and its absolute value is larger for the charged ligands.

As it follows from the results, taking into account electron correlation in the frame of MP2 (method QM5) is important for the accuracy of the scheme of the calculation. The extension of the basis set also allows to improve the coincidence of the $\Delta G_{b_calc}$ and $\Delta G_{b_exp}$.

As the τ value, correlation of the $\Delta G_{b_calc}$ and $\Delta G_{b_exp}$ can be also important especially in cases when it necessary to determine the relative stability of complexes. For the quantitative assessment of the correlation coefficient $\sigma$ can be used. The values of correlation the coefficient $\sigma$ are listed in Table 2. For each column $\sigma$ is calculated for the sum of the components of the $\Delta G_{b_calc}$ starting from the second to the given column inclusive. For example in the fourth column ($\Delta G_{vib_tr}$) the correlation coefficient is calculated between the $\Delta G_{b_exp}$ and the $\Delta E_0+\Delta G_{PCM}+\Delta G_{vib_tr}$. In the last column $\Delta E_{rel}$ is added to sum of the preceding components, which gives the complete $\Delta G_{b_calc}$ value.

As it shown from the Table 2, the highest level of correlation is achieved for the $\Delta E_0+\Delta G_{PCM}$. Addition of the next component $\Delta G_{vib_tr}$, improves agreement of the absolute values of the calculated and experimental free energies but decreases their correlation. The reason for this behavior can be as follows. Oscillation frequencies and corresponding contribution to $\Delta G_{b_calc}$ are computed in vacuum without taking into account the interaction between the explicit water molecules belonging to the complex and bulk water. The inclusion of the effect of the ligand transfer from the infinity to the second coordinating sphere ($\Delta G_{(a)}$) slightly improves the correlation.

Thus, the systematic calculations of relative stability of complexes formed by organic molecules with a $Mg^{2+}$ cation in water have been performed for the first time. The developed calculation procedure is applied to neutral molecules as well as to charged ones. It has been shown that the relaxation energy of water molecules should be taken into account (the term $\Delta E_{rel}$ in eq. (11)) to reproduce the experimental values of the complex formation free energy with the accuracy of 2 kkal/mol in calculations. The best correlation between experimental and calculated values of relative stability of complexes can be ensured by taking into account the coordinating bonds energy change and hydrogen bonds with water molecules rearrangement in the first coordinating sphere. These effects should be described in the frame of quantum chemical calculations with electron correlation ($\Delta E_0$) and continuum water PCM model ($\Delta G_{PCM}$) taken into account. To improve agreement with experiments the accuracy of calculations of the contribution from vibration degrees of freedom to the complex formation free energy should be improved.

## Conclusions

In the present work the methodology of calculation of formation free energy $\Delta G_b$ of organic molecule complexes with $Mg^{2+}$ cations in water solution have been derived. The employment of quantum chemical calculations with electron correlation taking into account and sufficiently rich basis sets with polarizable continuum implicit solvent model is the most important and time consuming stage of the calculations. Calculation of formation free energy $\Delta G_b$ for nine organic molecule complexes with $Mg^{2+}$ in water from NIST database [24] have been performed and compared with experimental results. It was shown that the main contribution to the $\Delta G_b$ is caused by the change of the complex internal energy resulting from the breaking/formation of coordination bonds and the change of the electrostatic interaction between the $Mg^{2+}$ ion and bulk water.

For all nine complexes of the organic molecules with a $Mg^{2+}$ cation taken from NIST database [16] the reasonable agreement (2 kcal/mol) of the calculated and experimental values of the $\Delta G_b$ was achieved.

**Acknowledgements**

This work supported by the Russian Foundation for Basic Research (grants no. 06-03-33171a, 07-02-01123a and 08-04-12129-ofi).

Table 1. The components of the free energy (in kcal/mol) (see (11)), calculated $\Delta G_{b_calc}$ and experimental $\Delta G_{b_exp}$ values of the energy of the complex formation, ligand charge Z (in au). Notations for the quantum chemical modeling levels are in the bottom of Table.

| | | I | II | III | IV | V | VI | VII | VIII | IX |
|---|---|---|---|---|---|---|---|---|---|---|
| $Z_L$ | | 0 | | | -1 | | | | | -2 |
| $\Delta E_0$ | QM1 | 1.0 | -15.6 | -14.8 | -15.7 | -24.4 | -20.1 | -16.9 | -20.9 | -56.0 |
| | QM5 | -3.2 | -14.4 | -14.3 | -16.9 | -25.2 | -20.7 | -17.8 | -21.3 | -53.3 |
| | QM3 | 4.2 | -16.0 | -14.4 | -14.5 | -23.6 | -20.2 | -18.1 | -22.7 | -51.1 |
| $\Delta G_{PCM}$ | QM1 | -0.6 | 8.0 | 6.6 | 11.7 | 17.3 | 10.2 | 3.8 | 8.4 | 40.1 |
| | QM5 | | | | | | | | | |
| | QM3 | -2.8 | 7.8 | 6.4 | 11.5 | 17.0 | 10.0 | 5.0 | 8.5 | 38.4 |
| $\Delta G_{vib_tr}$ | | 4.7 | 3.2 | 2.0 | 6.4 | 6.5 | 4.4 | 4.7 | 5.5 | 5.3 |
| $\Delta G_{cav}$ | | -0.4 | -0.2 | -0.1 | 0.3 | 0.2 | 0.1 | 0.1 | 0.1 | -0.1 |
| $\Delta G_{np}$ | | 0.8 | -0.3 | -0.3 | -0.7 | -0.6 | -0.1 | -0.1 | 0.0 | 0.2 |
| $\Delta E_{rel}$ | | 0.3 | 1.5 | 1.0 | 1.1 | 1.3 | 3.5 | 4.9 | 5.3 | 4.8 |
| $\Delta G_{b)}$ | | 1.0 | | | -1.0 | | | | | -3.1 |
| $\Delta G_{b_calc}$ | QM1 | 6.8 | -2.4 | -4.6 | 2.1 | -0.7 | -3.0 | -4.5 | -2.6 | -8.8 |
| | QM5 | 2.6 | -1.2 | -4.1 | 0.9 | -1.5 | -3.6 | -5.3 | -3.0 | -6.1 |
| | QM3 | 7.8 | -3.0 | -4.4 | 3.1 | -0.3 | -3.3 | -4.5 | -4.3 | -5.6 |
| $\Delta G_{b_exp}$ | | 0.5 | -0.7 | -2.6 | -0.1 | -3.9 | -5.1 | -6.5 | -6.4 | -4.5 |

QM1=B3LYP/6-31G**

QM2=MP2/6-31G**//B3LYP/6-31G**

QM3=B3LYP/cc-pVTZ//B3LYP/6-31G**

QM4=MP2/cc-pVTZ//B3LYP/6-31G**.

QM5=MP2/6-31G**

QM6=B3LYP/cc-pVTZ

Table 2. Values of correlation coefficient $\sigma_{QM5}$ for the quantum chemical modeling using the QM5 level between the $\Delta G_{b_calc}$ and the $\Delta G_{b_exp}$ value (explanation see the text).

| Comp. | $\Delta E_0$ | $\Delta G_{PCM}$ | $\Delta G_{vib_tr}$ | $\Delta G_{cav}$ | $\Delta G_{np}$ | $\Delta G_{(b)}$ | $\Delta G_{b_calc}$ |
|---|---|---|---|---|---|---|---|
| $\sigma_{QM5}$ | 0.46 | 0.84 | 0.76 | 0.75 | 0.75 | 0.79 | 0.80 |

Figure 1. Scheme of the second stage of $Mg^{2+}L$ complex formation reaction (L=pyridine). Molecules in the second CS are encircled.

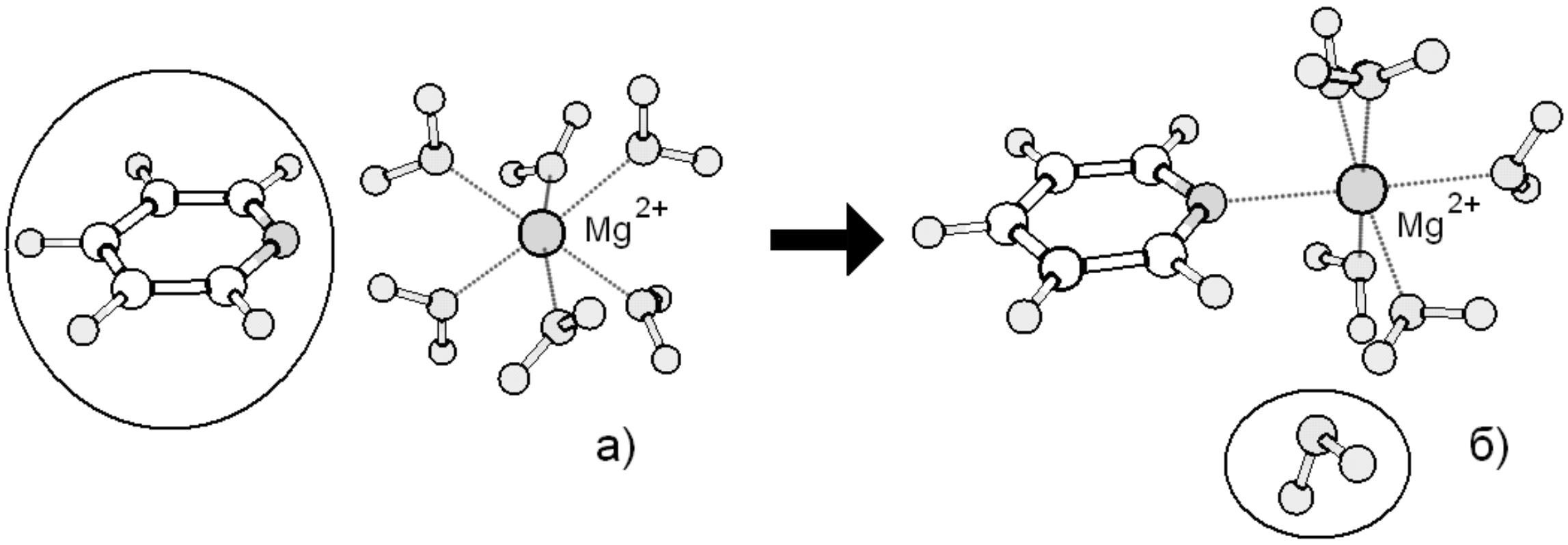

Figure 2. The structure of ligands examined to calculate the stability of chelate complexes with the magnesium ion.

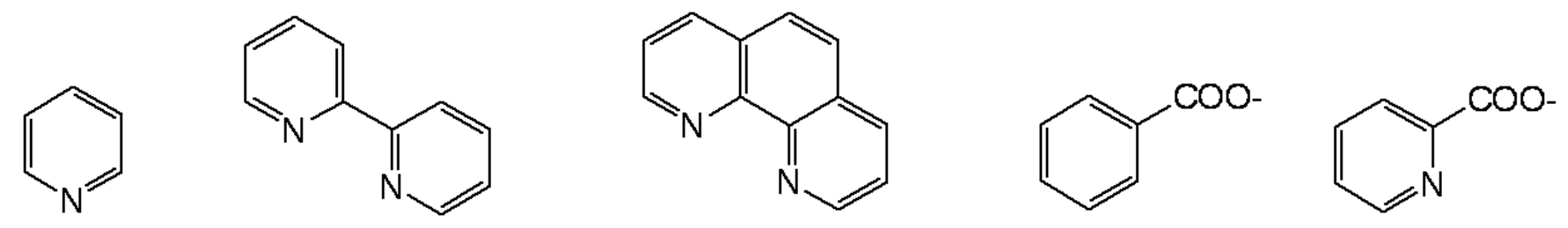

**pyridine (I)** **2,2'-bipyridine (II)** **1,10-phenanthroline (III)** **benzoic acid (IV)** **picolinic acid (V)**

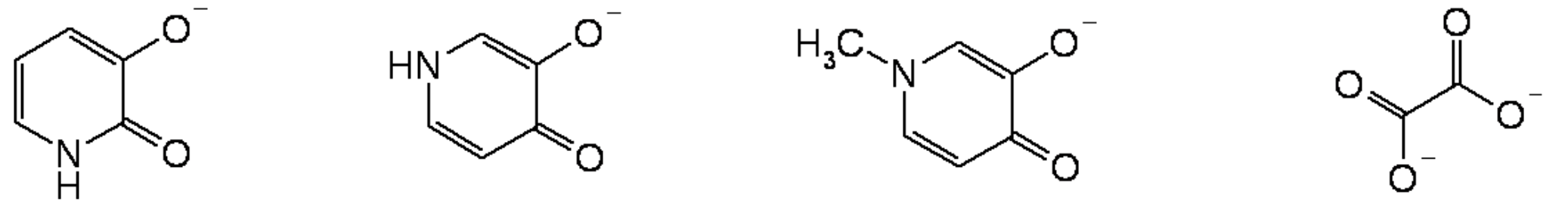

**3-hydroxypyridin-2(1*H*)-one (VI)** **3-hydroxypyridin-4(1*H*)-one (VII)** **3-hydroxy-1-methylpyridin-4(1*H*)-one (VIII)** **oxalic acid (IX)**